\documentclass[twocolumn,showpacs,showkeys,preprintnumbers,amsmath,amssymb]{revtex4}
\usepackage{graphicx}
\usepackage{dcolumn}
\usepackage{bm}

\begin{document}

\title{Systematic study of even-even $^{20-32}$Mg isotopes}

\author{Fouad~A.~Majeed}\email{fouadalajeeli@yahoo.com}
\affiliation{Department of Physics, College of Education for Pure
Science, University of Babylon, P.O.Box 4., Hilla-Babylon,Iraq}

\date{\today}

\begin{abstract}
A systematic study for 2$^+_1$ and 4$^+_1$ energies for even-even
$^{20-32}$Mg by means of large-scale shell model calculations
using the effective interaction USDB and USDBPN with SD and SDPN
model space respectively. The reduced transition probability  {\em
B}({\em E}2;$\uparrow$) were also calculated for the chain of Mg
isotopes.Very good agreement were obtained by comparing the first
2$^+_1$ and 4$^+_1$ levels for all isotopes with the recently
available experimental data and with the previous theoretical work
using 3DAMP+GCM model, but studying the transition strengths {\em
B}({\em E}2; 0$^+_{g.s.} \rightarrow$2$^+_1$) for Mg isotopes
using constant proton-neutron effective charges prove the
limitations of the present large-scale calculations to reproduce
the experiment in detail.
\end{abstract}

\maketitle

\section{Introduction}
The low energy structure of magnesium nuclei has attracted
considerable interests in the last decade, both experimental and
theoretical. In particular, the sequence of isotopes $^{20-40}$Mg
encompasses three spherical magic shell numbers : N=8, 20 and 28
and, therefore presents an excellent case for studies of the
evolution of shell structure with neutron number, weakening of
spherical shell closures, disappearance of magic numbers, and the
occurrence of {\em islands of inversion} \cite{OM08}.

 Extensive experimental studies of the low-energy structure of Mg isotopes
have been carried out at the Institute of Physical and Chemical
Research, Japan (RIKEN) \cite{H01,S09}, Michigan State University
(MSU) \cite{BV99,JM06,AL07,AG07} , the Grand Acc\'{e}l\'{e}rateur
National d'Ions Lourds, France (GANIL) \cite{VC01} and CERN
\cite{ON05,WS09}.

 In addition to numerous theoretical studies based on large-scale
shell-model calculations \cite{EC98,YU99,TO01,TT01,EC05, FM05},
the self-consistent mean-field framework, including the
nonrelativistic Hartree-Fock-Bogolibov (HFB) model with Skyrme
\cite{JT97} and Gogny forces \cite{RJ02} and the relativistic
mean-field (RMF) model \cite{SK91, ZZ96} as well as the
macroscopic-microscopic model based on a modified Nilsson
potential \cite{QJ06}, have been used to analyze the ground-state
properties (binding energies, charge radii, and deformations) and
low-lying excitation spectra of magnesium isotopes.

 The purpose of present work is to study the ground state 2$^+_1$ and
 4$^+_1$ excitation energies and the reduced transition probabilities {\em
B}({\em E}2; 0$^+_{g.s.}\rightarrow $2$^+_1$) (e$^{2}$fm$^{4}$) of
the even-even $^{20-32}$Mg isotopes using the new version of
Nushell@MSU for windows \cite{BW07} and compare these calculations
with the most recent experimental and theoretical work.

\section{Shell Model Calculations}
The calculations were carried out in the SD and SDPN model spaces
with the USDB and USDBPN effective interactions \cite{BW06} using
the shell model code  Nushell@MSU for windows \cite{BW07}.

The core was taken as $^{16}$O with 4 valence protons and
4,6,8,10,12,14,16 valence nucleons for $^{20}$Mg, $^{22}$Mg,
$^{24}$Mg, $^{26}$Mg, $^{28}$Mg, $^{30}$Mg and $^{32}$Mg
respectively distributed over 1{\em d}$_{5/2}$ , 1{\em d}$_{5/2}$
and 2{\em s}$_{1/2}$.

The effective interaction USDB with model space SD where used in
the calculation of the $^{20-30}$Mg isotopes,  while USDBPN in pn
formalism where employed with SDPN model space for $^{32}$Mg
nucleus.
\section{Results and Discussion}
The test of success of large-scale shell model calculations is the
predication of the low-lying 2$^+_1$ and 4$^+_1$ and the
transition rates {\em B}({\em E}2; 0$^+_{g.s.}
\rightarrow$2$^+_1$) using the optimized effective interactions
for the description of {\em sd}-shell nuclei.

Figure 1 presents the comparison of the calculated {\em
E}$_x$(2${^+_1}$) energies from the present work (P.W.) with the
experiment \cite{NN12}, the work of J. M. Yao {\em et al.}\cite{JM11} using 3DAMP+GCM model with the
relativistic density functional PC-F1.

The comparison shows that our calculation are in better agreement with the experiment than the work of Ref.\cite{JM11}
\begin{figure}
\centering
\includegraphics[width=0.44\textwidth]{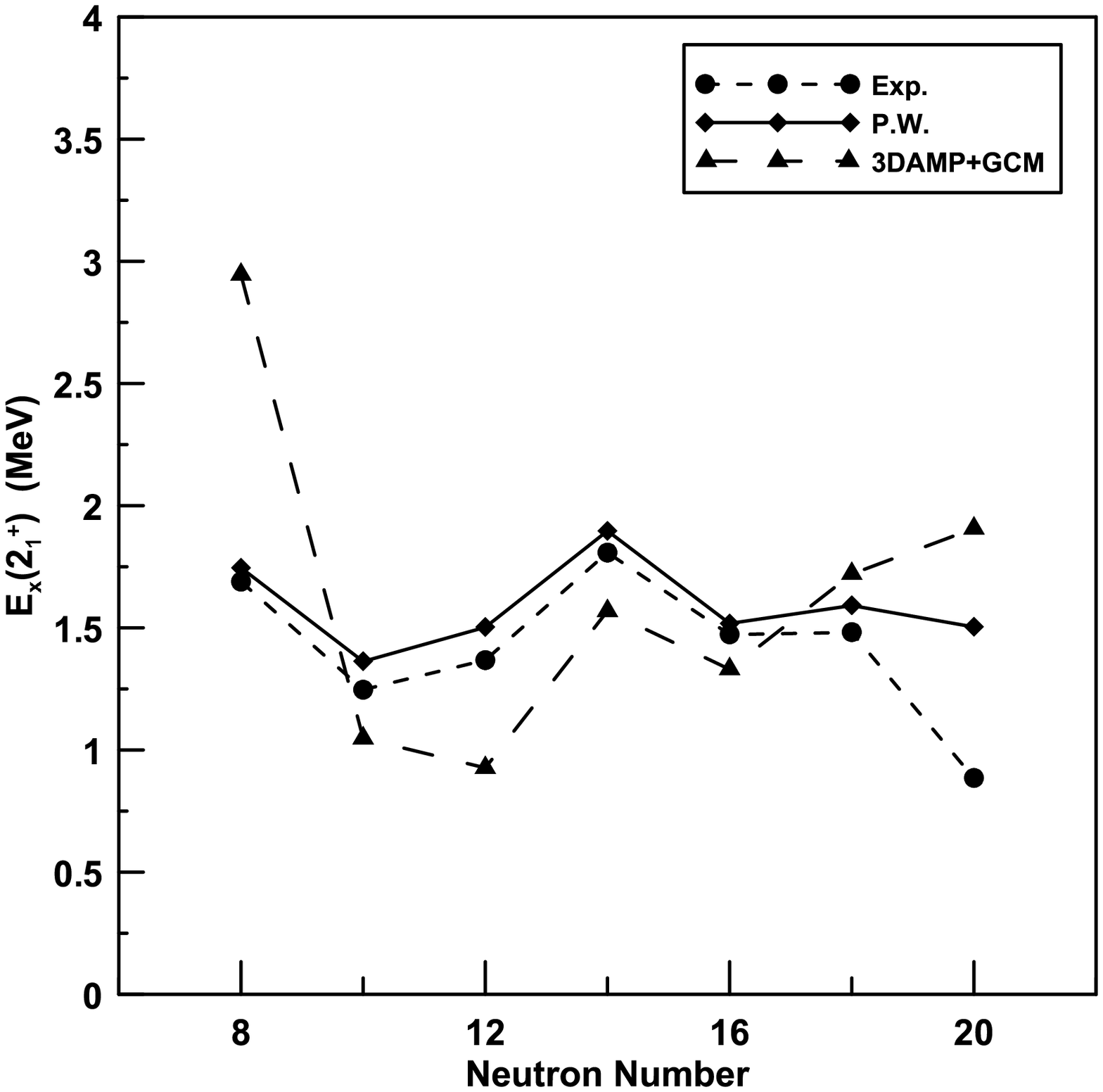}
\caption{Systematics of {\em E}$_x$(2${^+_1}$) for eve-even Mg
isotopes. Experimental data (closed circles) are compared with
present work (solid line), the previous work using 3DAMP+GCM model
(long dashed line). Experimental data are
taken from Ref.\cite{NN12}.}
\end{figure}

Figure 2 shows the comparison of the calculated low-lying {\em
E}$_x$(4${^+_1}$) excitation energies from present work (P.W.)with the
experiment \cite{NN12}, the work of J. M. Yao {\em et al.}\cite{JM11} using 3DAMP+GCM model with the
relativistic density functional PC-F1.
The comparison shows very clear that our prediction for the {\em
E}$_x$(4${^+_1}$) are in better agreement with the experiment.

Figure 3 presents the comparison of the calculated
{\em B}({\em E}2; 0$^+_{g.s.}\rightarrow $2$^+_1$) (e$^{2}$fm$^{4}$)
from present work (P.W.) with the experimental data taken from the Institute of Physical and Chemical
Research, Japan (RIKEN) \cite{H01,S09}, the Grand Acc\'{e}l\'{e}rateur
National d'Ions Lourds, France (GANIL) \cite{VC01} and CERN
\cite{ON05,WS09}, the previous theoretical work of J. M. Yao {\em et al.}\cite{JM11} using 3DAMP+GCM model
and with the work of R. Rodr\'{i}guez-Guzm\'{a}n {\em et al.}\cite{RJ02} using HFB-Gogny force.
The effective charges were taken to be e$_{\pi}$=1.25e for proton and e$_{\nu}$=0.8e for neutron. With these effective charges
our prediction for the reduced transition probability {\em B}({\em E}2; 0$^+_{g.s.}\rightarrow $2$^+_1$) are more closer to the
experimental values than the previous work of Refs. \cite{JM11,RJ02}.

\begin{figure}
\centering
\includegraphics[width=0.44\textwidth]{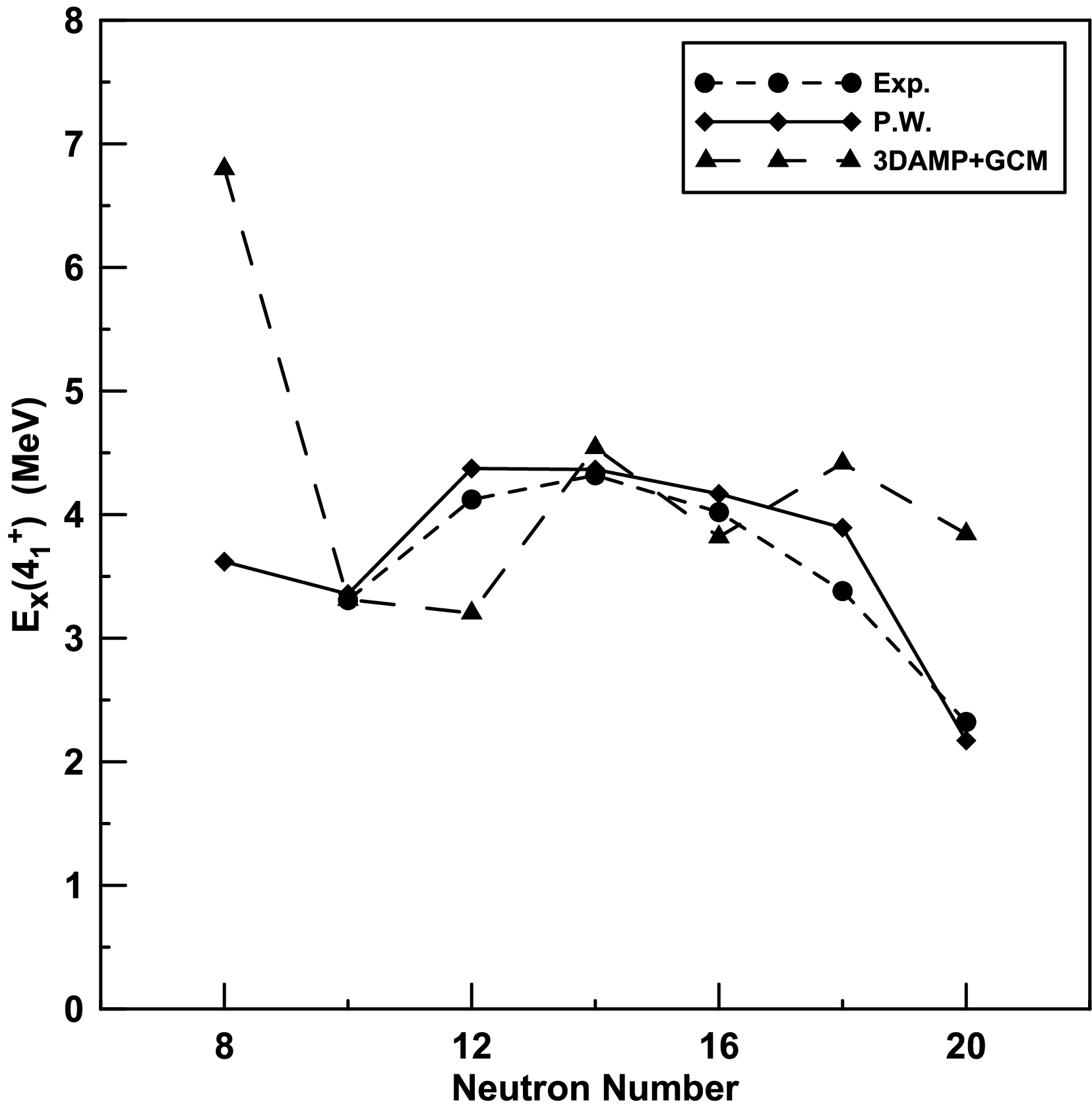}
\caption{Systematics of {\em E}$_x$(4${^+_1}$) for eve-even Mg
isotopes. Experimental data (closed circles) are compared with
present work (solid line), the previous work using 3DAMP+GCM model
(long dashed line). Experimental data are
taken from Ref.\cite{NN12}.}
\end{figure}

\begin{figure}
\centering
\includegraphics[width=0.5\textwidth]{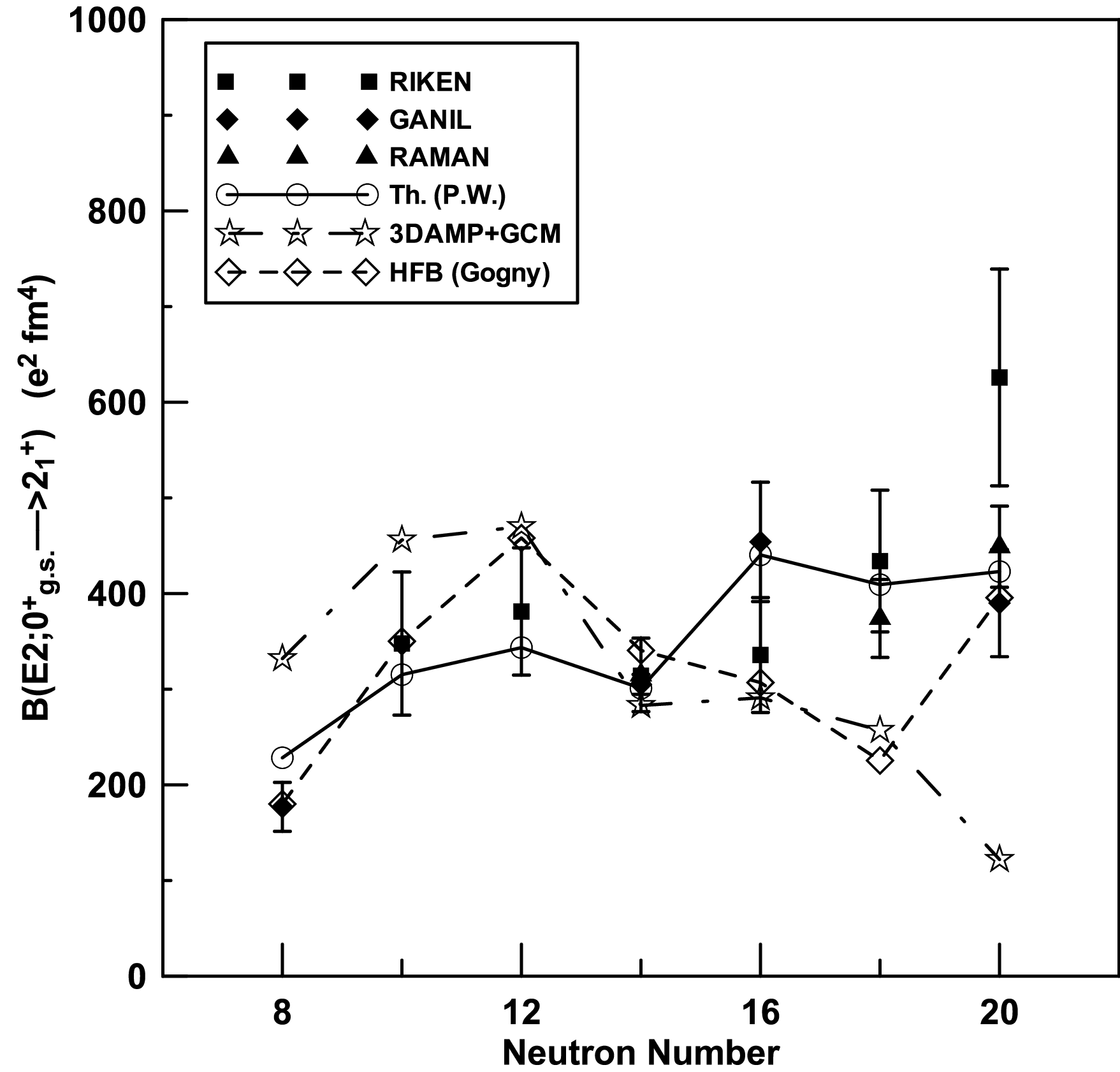}
\caption{Comparisons between the calculated {\em
B}({\em E}2; 0$^+_{g.s.}\rightarrow $2$^+_1$) (e$^{2}$fm$^{4}$) of
the even-even $^{20-32}$Mg isotopes (solid line) (P.W.), 3DAMP+GCM model (dashed-dotted line)\cite{JM11}
and HFB (Gogny)(dashed line) \cite{RJ02}. Experimental data taken from Refs. \cite{H01,S09,VC01,ON05,WS09}}
\end{figure}
\newpage
\section{Summary}
Unrestricted large scale-shell model calculations were performed using the effective interactions USDB and USDBPN
in pn formalism with the model space SD and SDPN to study the low lying 2$^+_1$ and 4$^+_1$ energies for even-even
$^{20-32}$Mg isotopes and the transition strengths {\em B}({\em E}2; 0$^+_{g.s.}
\rightarrow$2$^+_1$) for the mass region A=20-32.
Good agreement were obtained in comparing our theoretical work with
the recent available experimental data and with the most recent theoretical work of Ref.\cite{JM11} using 3DAMP+GCM model with the
relativistic density functional PC-F1 for both excitation energies and transition strengths.
\newpage


\end{document}